\documentclass[preprint,12pt]{elsarticle}

\usepackage{graphicx}

\usepackage{amssymb, amsthm}

\biboptions{compress}

\journal{\hspace{-3.5cm} \raisebox{-1mm}{\begin{tikzpicture} \draw [fill = white, white] (0, 0) rectangle (3.5, 0.4); \end{tikzpicture}}}

\usepackage{amsmath}
\usepackage{url}
\usepackage{color}
\usepackage{tensor}
\usepackage{mathrsfs}
\usepackage{nicefrac}
\usepackage{tikz-cd}
\usepackage{tikz}
\usetikzlibrary{decorations.markings}
\usetikzlibrary{matrix,arrows}
\usepackage{array}
\usepackage{ulem}
\usepackage{placeins}

\usepackage{pdflscape}

\usepackage[latin1]{inputenc}      
\usepackage[T1]{fontenc}    
\usepackage{ae,aecompl}                          
\usepackage{dsfont}
\usepackage{appendix}
\usepackage{amsfonts,amsmath,latexsym,bbm}
\usepackage{amssymb}
\usepackage{accents}
\usepackage{graphics}
\usepackage{graphicx}
\usepackage{booktabs,longtable}
\usepackage{bm}
\usepackage{dcolumn}
\usepackage{enumitem}
\usepackage{upgreek}
\usepackage{tensor}

\usepackage{tikz}
\usepackage{tikz-cd}
\usepackage{tikz-3dplot}
\usetikzlibrary{decorations.markings}
\usetikzlibrary{matrix,arrows}
\usetikzlibrary{calc}
\usetikzlibrary{shapes}
\usetikzlibrary{positioning}
\usetikzlibrary{fit}

\usepackage{pgfplots}
\pgfplotsset{width=7cm,compat=1.5.1}

\usetikzlibrary{decorations.pathreplacing}
\makeatletter
\def\underbrace#1{%
  \@ifnextchar_{\tikz@@underbrace{#1}}{\tikz@@underbrace{#1}_{}}}
\def\tikz@@underbrace#1_#2{%
  \tikz[baseline=(a.base)] {\node[inner sep=2] (a) {\(#1\)};
  \draw[thick,line cap=round,decorate,decoration={brace,amplitude=4pt}]
    (a.south east) -- node[pos=0.5,below,inner sep=7pt] {\(\scriptstyle #2\)} (a.south west);}}
\makeatother

\usepackage[linktocpage=true]{hyperref}
\usepackage{hyperref}

\makeatletter
\renewcommand*{\eqref}[1]{%
  \hyperref[{#1}]{\textup{\tagform@{\ref*{#1}}}}%
}
\makeatother

\setlist{font=\normalfont\itshape} 

\renewcommand{\vec}[1]{\boldsymbol{#1}}

\newcommand{\h}{\hspace{1pt}}
\newcommand{\hh}{\hspace{0.5pt}}
\newcommand{\mh}{\hspace{-1pt}}

\renewcommand{\i}{\mathrm i}

\definecolor{oldgray}{gray}{0.4}

\DeclareMathAlphabet{\mathbbmsl}{U}{bbm}{m}{sl}

\numberwithin{equation}{section}

\begin{document}

\begin{frontmatter}



\title{Why history matters: {\itshape ab initio} rederivation of Fresnel \mbox{\hspace{-2pt}equations confirms microscopic theory of refractive index}}


\author[freiberg]{R.~Starke}
\ead{Ronald.Starke@physik.tu-freiberg.de}

\author[heidelberg,aachen]{G.A.H.~Schober\corref{cor1}}
\ead{schober@physik.rwth-aachen.de}

\cortext[cor1]{Corresponding author.}

\address[freiberg]{Institute for Theoretical Physics, TU Bergakademie Freiberg, \\ Leipziger Stra\ss e 23, 09596 Freiberg, Germany \vspace{0.1cm}}
\address[heidelberg]{Institute for Theoretical Physics, Heidelberg University, \\ Philosophenweg 19, 69120 Heidelberg, Germany \vspace{0.1cm}}
\address[aachen]{Institute for Theoretical Solid State Physics, RWTH Aachen University, Otto-Blumenthal-Stra\ss e 26, 52074 Aachen}

\begin{abstract}
We provide a systematic theoretical, experimental, and historical critique of the standard derivation of Fresnel's equations, which shows in particular that these  well-established equations actually contradict the traditional, macroscopic approach to electrodynamics in media. Subsequently, we give a rederi\-{}vation of Fresnel's equations which is exclusively
based on the microscopic Maxwell equations and hence in accordance with modern first-principles materials physics. In particular, as a main outcome of this analysis being of a more general interest, we propose the most general boundary conditions on electric and magnetic fields
which are valid on the microscopic level.
\end{abstract}




\end{frontmatter}



\newpage
\tableofcontents

\newpage
\section{Introduction}\label{sec:intro}

Introduced by P.~Nozi\`{e}res and D.~Pines in 1958 as the ``generalized dielectric constant at arbitrary
frequency $\omega$ and wavevector $\vec k$'' \cite[p.~470, notation adapted]{NozieresPines2}, the {\it dielectric function} soon turned into a concept of outstanding importance 
for {\it ab initio} materials physics.
On the theoretical side, it had already been instrumental for the very development of first-principles methods 
\cite{NozieresPines,Lindhard,Adler,Wiser,Ehrenreichcohen}, only to become itself a prime target quantity for both Green function theory and density functional theory (see e.g.~Refs.~\cite{Hanke,Strinati,Onida} for classical reviews).
Practically, knowledge of the dielectric function does not only allow one to access---via the zero-frequency limit---the age-old {\it dielectric constant}, 
but also the {\it optical conductivity}, the {\it density response function}, and finally---via the famous Maxwell relation---even the {\it refractive index}
and its ensuing optical properties such as the {\it reflectance}. At present, the prediction of the dielectric function therefore
ranks among the most active areas in both experimental and theoretical research (see Refs.~\cite{Trevisanutto13,LiLi,Yang15,Arbodela16,Bejaoui16,Feneberg16,Hassanien16diel,Nuzhnyy16,Seyidov16,Vor17,Zheng17} for recent examples).

However, as has become clear since the advent of the Modern Theory of Polarization \cite{Resta07,Resta10,Vanderbilt}, 
current first-principles methods (see e.g.~Refs. \cite{Giuliani, Kohanoff,Martin,Bechstedt,SchafWegener,Kaxiras,MartinRothen} for modern textbooks) used to access the dielectric function   
differ profoundly from their traditional counterparts.
While the latter had been formulated {\it macroscopically} in terms of ``dipole densities'', ab initio methods rely
on a {\it microscopic} formulation in terms of ``external'' and ``induced'' electromagnetic~fields. In fact, this provides a new framework of electrodynamics in media, which is in stark contrast
to most traditional textbooks (see e.g.~Refs.~\cite{Jackson,Griffiths,Landau}).
Recently, the authors of this article have therefore condensed the corresponding common practice of ab initio materials physics into
the Functional Approach to electrodynamics of media \cite{ED1,ED2,EDOhm,Refr,EDLor,EDWave,EDFullGF,EffWW}, which is an inherently microscopic theory of electromagnetic material properties. 
With these developments, however, the following fundamental problem arises: the classical methods for the {\itshape measurement} of the dielectric function, 
such as ellipsometry or reflectivity spectroscopy (see e.g.~Ref.~\cite[\S\,6.1.2]{Cardona}), rely on {\it Fresnel's equations,}
and these are usually derived in the phenomenological framework of macroscopic electrodynamics (see e.g.~Refs.~\cite[\S\,7.3]{Jackson} or \cite[\S\,9.3.3]{Griffiths}).
It therefore remains to prove that Fresnel's equations can also be justified within the microscopic framework
which is actually used for the calculation of 
the dielectric function from first principles. This is precisely the main objective of the present article.

Concretely, we address this problem as follows. After a short introduction to Fresnel's equations in Sct.~\ref{Sec:FresnelEqus}
and a review of their standard derivation in \S\,\ref{subsec:standardDeriv}, we proceed in \S\,\ref{subsec:histReview} with a historical review of the experimental and theoretical devolopments
which eventually led to this standard derivation. With this, we go on to subject the standard derivation of Fresnel's equations to a thorough critique (\S\,\ref{subsec:critique}).
Finally, in Sct.~\ref{subsec:derivation}, we discuss the problem of {\itshape general} electromagnetic boundary conditions on the microscopic level, and we present a completely {\it microscopic} derivation of Fresnel's equations.

\section{Fresnel equations and optical properties} \label{Sec:FresnelEqus}

We consider an ordinary refraction experiment with an incident light ray impinging on the flat interface between two materials with refractive indices $n_1$ and $n_2$\hh.
In most cases, the incident ray will be split into a reflected part and a transmitted (or refracted) part, whose directions are given in terms of the respective refractive indices by the {\itshape law of reflection,}
\begin{equation}
 \varphi_{\rm r}=\varphi_1 \,,
\end{equation}
and by {\itshape Snellius' law of refraction,}
\begin{equation}
n_1\sin\varphi_1=n_2\sin\varphi_2\,. \label{eq_lawRefr}
\end{equation}
Here, $\varphi_1$ is the angle of incidence measured with respect to the surface normal, while $\varphi_{\rm r}$ and $\varphi_2$ are the respective angles of the reflected and the refracted light rays.
In this situation, the {\it Fresnel equations} determine the {\it polarization-dependent} electric field amplitudes 
of the reflected and the transmitted rays relative to the field amplitude of the incident ray. Concretely, Fresnel's equations read as follows (see, for example, Ref.~\cite[Eqs.~(4.270)--(4.273)]{NoltingEdyn}):
\begin{align}
\left(\frac{E_{\rm t}}{E_{\rm i}}\right)_{\!\!\rm s}&=\frac{2 \h n_1\cos\varphi_1}{n_1\cos\varphi_1+n_2\cos\varphi_2}\,,\label{eq_Fresnel1}\\[5pt]
\left(\frac{E_{\rm r}}{E_{\rm i}}\right)_{\!\!\rm s}&=\frac{n_1\cos\varphi_1-n_2\cos\varphi_2}{n_1\cos\varphi_1+n_2\cos\varphi_2}\,,\label{eq_Fresnel2}\\[5pt]
\left(\frac{E_{\rm t}}{E_{\rm i}}\right)_{\!\!\rm p}&=\frac{2 \h n_1\cos\varphi_1}{n_2\cos\varphi_1+n_1\cos\varphi_2}\,,\label{eq_Fresnel3}\\[5pt]
\left(\frac{E_{\rm r}}{E_{\rm i}}\right)_{\!\!\rm p}&=\frac{n_2\cos\varphi_1-n_1\cos\varphi_2}{n_2\cos\varphi_1+n_1\cos\varphi_2}\,.\label{eq_Fresnel4}
\end{align}
Here, $E_{\rm i}$, $E_{\rm r}$ and $E_{\rm t}$ respectively denote the electric field amplitudes of the incident, reflected and transmitted (or refracted) light rays.
Finally, the subscripts ``s'' and ``p'' refer to the polarization orthogonal (``senkrecht'') and parallel to the plane of incidence (which is defined by
the direction of the incident ray and the surface normal). By invoking Snellius' law of refraction,
the above equations further simplify to \cite[Eqs.~(4.274)--(4.277)]{NoltingEdyn}
\begin{align}
\left(\frac{E_{\rm t}}{E_{\rm i}}\right)_{\!\!\rm s}&=\frac{2 \hh \cos\varphi_1\sin\varphi_2}{\sin(\varphi_2+\varphi_1)}\,, \label{orig_Fresnel1} \\[5pt]
\left(\frac{E_{\rm r}}{E_{\rm i}}\right)_{\!\!\rm s}&=\frac{\sin(\varphi_2-\varphi_1)}{\sin(\varphi_2+\varphi_1)}\,,\label{eq_FresnelSineLaw}\\[5pt]
\left(\frac{E_{\rm t}}{E_{\rm i}}\right)_{\!\!\rm p}&=\frac{2 \hh \cos\varphi_1\sin\varphi_2}{\sin\varphi_1\cos\varphi_1+\sin\varphi_2\cos\varphi_2}\,, \label{orig_Fresnsel3} \\[5pt]
\left(\frac{E_{\rm r}}{E_{\rm i}}\right)_{\!\!\rm p}&=\frac{\tan(\varphi_1-\varphi_2)}{\tan(\varphi_1+\varphi_2)}\,.\label{eq_FresnelTangentLaw}
\end{align}
These are in fact the original equations due to Augustin-Jean Fresnel (1788-1827). Correspondingly, Eqs.~\eqref{eq_FresnelSineLaw} and \eqref{eq_FresnelTangentLaw}
are known as ``Fresnel's sine law'' and ``Fresnel's tangent law'', respectively. Furthermore, the ratios $r_{\rm s} := (E_{\rm r}/E_{\rm i})_{\rm s}$ and $r_{\rm p} := (E_{\rm r}/E_{\rm i})_{\rm p}$ of the reflected field amplitudes to their incident counterpart  are called {\it reflectivities}.

Apart from their outstanding theoretical value to be clarified below, the Fresnel equations are also of tremendous practical importance. Let us mention only the most 
important applications: 

\begin{enumerate}
 \item[(i)]{\itshape Brewster angle.}
Setting Eq.~\eqref{eq_Fresnel4} to zero leads to the condition
\begin{equation}
n_2\cos\varphi_1=n_1\cos\varphi_2\,.
\end{equation}
Eliminating form this the variable $\varphi_2$ via the law of refraction, Eq. \eqref{eq_lawRefr}, we obtain after some algebra the equation \cite[Eq.~(4.278)]{NoltingEdyn}
\begin{equation}
\tan\varphi_{1} =\frac{n_2}{n_1}\,,  \label{eq_Brewster}
\end{equation}
which is the formula for the {\itshape Brewster angle.} For light incident under this angle, the intensity of the reflected light with polarization 
parallel to the plane of incidence vanishes.

\item[(ii)]{\itshape Reflectance.}
In vacuo, the ratio between incident and reflected intensities, i.e., the {\itshape reflectance,} is given by the squared modulus of the  ratio between the corresponding field amplitudes (i.e., of the reflectivity). Concretely, considering normal incidence from the vacuum ($n_1=1$), Snellius' law implies that the angles of incidence and reflection are equal, hence $\varphi_1=\varphi_2=0$.
With this, it is straightforward to show that the reflectance is independent of the polarization, \smallskip
\begin{equation}
 R=|r_{\rm s}|^2=|r_{\rm p}|^2 \,, \smallskip
\end{equation}
which is intuitive since for normal incidence the distinction between s- and p-polarization looses its meaning anyway. Furthermore, the reflectance is given by (see e.g.~Refs.~\cite[Eq.~(6.8)]{Cardona} or \cite[Eq.~(1.29)]{Fox})
\begin{equation}
R=\left|\frac{n_2 - 1}{n_2 + 1}\right|^{2} = \frac{(\nu-1)^2+\kappa^2}{(\nu+1)^2+\kappa^2}\,, \label{eq_basisReflSpect}
\end{equation}
where $\nu:=\mathfrak{Re}\,\h n_2$ and $\kappa:=\mathfrak{Im}\,\h n_2$ are the so-called {\it optical constants,} with $\kappa$ being the {\it extinction coefficient}.

\item[(iii)]{\itshape Ellipsometry.} 
Finally, in ellipsometry experiments one considers again linearly polarized light incident from vacuum $(n_1 = 1)$ but at a non-zero angle $\varphi_1$. 
In general, the reflected ray is then elliptically polarized, and from the Fresnel equations it can be shown directly that 
\begin{equation}
(\nu+\i \kappa)^2 = \sin^2 \! \varphi_1 + \sin^2\!\varphi_1 \hh \tan^2 \! \varphi_1 \left(\frac{1-r_{\rm p}/r_{\rm s}}{1+r_{\rm p}/r_{\rm s}} \right)^{\!\mh 2}\,, \label{eq_ellipso}
\end{equation}
from which the refractive index can be computed in terms of the reflectivities (see e.g.~Ref.~\cite[Eq.~(6.13)]{Cardona}). As the ratio of the reflectivities is a measurable quantity, 
Eq.~\eqref{eq_ellipso} forms the basis of ellipsometric measurements of the dielectric function via the standard relation \mbox{$n^2=\varepsilon_{\rm r}$\hh.}
We note, in particular, that this would not be possible if the refractive index was given by the formula $n^2=\varepsilon_{\rm r}\h\mu_{\rm r}$\hh.
\end{enumerate}

\vspace{2pt}\noindent
Apart from ellipsometric measurements, Eqs.~\eqref{eq_Brewster} and \eqref{eq_basisReflSpect} are nowadays the basis for the deduction of the refractive index from the Brewster angle and from reflectivity spectra, respectively.

\section{Standard Approach}

\subsection{Standard derivation of Fresnel equations}\label{subsec:standardDeriv}

In the Standard Approach to electrodynamics in media, the Fresnel equations are derived from the so-called boundary conditions on the electromagnetic
fields, which read (see e.g.~Ref.~\cite[Eqs.~(I.17)--(I.20)]{Jackson})
\begin{align}
\vec n\cdot(\vec D_2-\vec D_1)&=\rho_{\partial V}\,, \label{eq_BCstand1}\\[3pt]
\vec n\times(\vec E_2-\vec E_1) &= 0\,, \label{eq_BCstand2}\\[3pt]
\vec n\cdot(\vec B_2-\vec B_1)&= 0\,, \label{eq_BCstand3}\\[3pt]
\vec n\times(\vec H_2-\vec H_1)&=\vec j_{\partial V} \,, \label{eq_BCstand4}
\end{align}
where $\rho_{\partial V}$ and $\vec j_{\partial V}$ are the surface charge and current densities, and $\vec n$ denotes the surface normal 
pointing from material ``1'' to material ``2''. To obtain the Fresnel equations, one then has to make three fundamental assumptions:
\begin{enumerate}
 \item The incident, reflected and transmitted light rays are represented by transverse electromagnetic plane waves, with their corresponding wave\-{}vectors being identical to the respective ray directions.
 \item The surface currents and charges are set to zero:
\begin{align}
 \rho_{\partial V} & = 0 \,, \\[3pt]
 \vec j_{\partial V}& = 0 \,.
\end{align}
 \item The fields $\vec D_j$ and $\vec H_j$ are eliminated by the conventional material relations (for $j = 1, 2$),
\begin{align}
 \vec D_{j} & =\varepsilon_0 \h\hh \varepsilon_{{\rm r}, \hh j} \h\hh \vec E_{j} \,, \\[5pt]
 \vec H_{j} & =(\mu_0 \h\hh \mu_{{\rm r}, \hh j})^{-1} \h \vec B_{j} \,, 
\end{align}
where the respective material parameters $\varepsilon_{{\rm r}, \hh j}$ and $\mu_{{\rm r}, \hh j}$ are assumed to be frequency-dependent constants.
\end{enumerate}

\smallskip \noindent
In particular, the second and the third assumption together with Eqs.~\eqref{eq_BCstand1}--\eqref{eq_BCstand4} imply the following boundary conditions for the electric field and the magnetic field (see e.g.~Ref.~\cite[Eqs.~(7.37)]{Jackson}):
\begin{align}
\vec n\cdot(\varepsilon_{\rm r, \hh 2} \h \vec E_2-\varepsilon_{\rm r, \hh 1} \h \vec E_1)&=0\,, \label{eq_BCrepl1}\\[3pt]
\vec n\times(\vec E_2-\vec E_1) &= 0\,, \label{eq_BCrepl2}\\[3pt]
\vec n\cdot(\vec B_2-\vec B_1)&= 0\,, \label{eq_BCrepl3}\\[3pt]
\vec n\times(\vec B_2/\mu_{\rm r, \hh 2}-\vec B_1/\mu_{\rm r, \hh 1})&=0\,. \label{eq_BCrepl4}
\end{align}
On account of the first assumption, one now starts from the following ansatz: the respective light rays are given by the Fourier-mode contributions
\begin{align}
\vec E_{\rm i}(\vec x,t) &= \vec E_{\rm i, 0}\h\exp(-{\rm i} \hh \omega_{\rm i} \hh t + {\rm i} \hh \vec k_{\rm i}\cdot\vec x)\,,\\[3pt]
\vec E_{\rm r}(\vec x,t) &= \vec E_{\rm r, 0}\h\exp(-{\rm i} \hh \omega_{\rm r} \hh t + {\rm i} \hh \vec k_{\rm r}\cdot\vec x)\,,\\[3pt]
\vec E_{\rm t}(\vec x,t) &= \vec E_{\rm t, 0}\h\exp(-{\rm i} \hh \omega_{\rm t} \hh t + {\rm i} \hh \vec k_{\rm t}\cdot\vec x)\,,
\end{align}
where the plane waves are assumed to be transverse, i.e.,
\begin{equation}
\vec k_{\rm i} \cdot\vec E_{\rm i} \h = \h \vec k_{\rm r} \cdot\vec E_{\rm r} \h = \h \vec k_{\rm t} \cdot\vec E_{\rm t}  = 0\,,
\end{equation}
and the magnetic fields are related to the electric fields via Faraday's law,
\begin{align}
\vec B_{\rm i} & = \vec k_{\rm i} \times \vec E_{\rm i} \hh / \omega_{\rm i} \,, \\[3pt]
\vec B_{\rm r} & = \vec k_{\rm r} \times \vec E_{\rm r} \hh / \omega_{\rm r} \,, \\[3pt]
\vec B_{\rm t} & = \vec k_{\rm t} \times \vec E_{\rm t} \hh / \omega_{\rm t} \,.
\end{align}
Furthermore, the frequencies are related to the wavevectors via the respective refractive indices, hence
\begin{equation}
\frac{\omega_{\rm i}}{c \hh |\vec k_{\rm i}|} = \frac{\omega_{\rm r}}{c \hh |\vec k_{\rm r}|} = \frac{1}{n_1} \,, \qquad \frac{\omega_{\rm t}}{c \hh | \vec k_{\rm t}|} = \frac{1}{n_2} \,.
\end{equation}
Next, the amplitudes $\vec E_{\rm i, 0}$\hh, $\vec E_{\rm r, 0}$ and $\vec E_{\rm t, 0}$ have to be chosen such that 
the electric and magnetic fields match the boundary conditions \eqref{eq_BCrepl1}-\eqref{eq_BCrepl4} with
\begin{align}
\vec E_1&:=\vec E_{\rm i}+\vec E_{\rm r}\,, \label{eq_E_in_refl}\\[3pt]
\vec E_2&:=\vec E_{\rm t}\,,\label{eq_E_trans}
\end{align}
and analogous equations for the magnetic fields.
Independently of the polarization, these boundary conditions then imply the constancy of the frequency,
\begin{equation}\label{eq_constantFrequ}
\omega_{\rm i}=\omega_{\rm r}=\omega_{\rm t} \h \equiv \h \omega \,, 
\end{equation}
and moreover, they even re-imply the laws of reflection and refraction, which can then be condensed into (see Ref.~\cite[Eq.~(4.6)]{Hecht})
\begin{equation}
\vec n\times\vec k_{\rm i} = \vec n\times\vec k_{\rm r} = \vec n\times\vec k_{\rm t}\,.
\end{equation}
Furthermore, for the ratios of the field amplitudes one finds after a more 
involved calculation (see Ref.~\cite[Eqs.~(4.262), (4.264), (4.267) and (4.268)]{NoltingEdyn}):
\begin{align}
\left(\frac{E_{\rm t}}{E_{\rm i}}\right)_{\!\!\rm s}&\stackrel{?}{=}\h\frac{2 \h Z^{-1}_1\cos\varphi_1}{Z^{-1}_1\cos\varphi_1+Z^{-1}_2\cos\varphi_2}\,,\label{eq_pseudoFresnel1}\\[5pt]
\left(\frac{E_{\rm r}}{E_{\rm i}}\right)_{\!\!\rm s}&\stackrel{?}{=}\h\frac{Z^{-1}_1\cos\varphi_1-Z^{-1}_2\cos\varphi_2}{Z^{-1}_1\cos\varphi_1+Z^{-1}_2\cos\varphi_2}\,,\label{eq_pseudoFresnel2}\\[5pt]
\left(\frac{E_{\rm t}}{E_{\rm i}}\right)_{\!\!\rm p}&\stackrel{?}{=}\h\frac{2 \h Z^{-1}_1\cos\varphi_1}{Z^{-1}_2\cos\varphi_1+Z^{-1}_1\cos\varphi_2}\,,\label{eq_pseudoFresnel3}\\[5pt]
\left(\frac{E_{\rm r}}{E_{\rm i}}\right)_{\!\!\rm p}&\stackrel{?}{=}\h\frac{Z^{-1}_2\cos\varphi_1-Z^{-1}_1\cos\varphi_2}{Z^{-1}_2\cos\varphi_1+Z^{-1}_1\cos\varphi_2}\,,\label{eq_pseudoFresnel4}
\end{align}
where $E = |\vec E|$\hh, etc., and where $Z$ denotes the {\it wave impedance} defined for each material as
\begin{align}
Z:=\sqrt{\frac{\mu_{\rm r}}{\varepsilon_{\rm r}}}\,.
\end{align}
Clearly, these formulae do not reproduce the Fresnel equations \eqref{eq_Fresnel1}--\eqref{eq_Fresnel4} if the refractive index is given by the standard formula
\begin{equation}
n\stackrel{?}{=}\sqrt{\varepsilon_{\rm r} \h \mu_{\rm r}}\,.
\end{equation}
We will therefore refer to Eqs.~\eqref{eq_pseudoFresnel1}--\eqref{eq_pseudoFresnel4} as the {\it pseudo-Fresnel equations}.
In order to reproduce the real Fresnel formulae from these, 
one has to introduce the following {\it additional assumption} (see e.g.~Ref.~\cite[\S\,7.3]{Jackson}):
\begin{enumerate} \setcounter{enumi}{3}
 \item The relative permeabilities of both materials are equal to one, i.e.,
\begin{align}
  \mu_{\rm r, \hh 1} = \mu_{\rm r, \hh 2} = 1  \,.
\end{align}
\end{enumerate}

\smallskip \noindent
With this additional assumption, it follows that
\begin{equation}
n_j=\sqrt{\varepsilon_{{\rm r}, \hh j}} \,,
\end{equation}
and hence, \smallskip
\begin{equation}
\frac{1}{Z_j} = n_j \,. \smallskip
\end{equation}
In this case, Eqs.~\eqref{eq_pseudoFresnel1}--\eqref{eq_pseudoFresnel4} imply the original Fresnel equations \eqref{eq_Fresnel1}--\eqref{eq_Fresnel4}, and this completes their derivation in the Standard Approach.

\subsection{Historical review}\label{subsec:histReview}

Both to put the above considerations into perspective and for the critique~of the Standard Approach
to be spelled out in the following subsection, we now provide a short historical account of the events that have finally led to the derivation of the Fresnel equations in the Standard Approach. We first summarize the developments in experimental physics up to the discovery of the Fresnel equations (\S\,\ref{experiment}), then shortly review the corresponding developments in theoretical physics (\S\,\ref{theory}), and finally conclude with a comparative discussion of these events in \S\,\ref{hist_conc}.
For short and readable introductions to the history of optics, the interested
reader is referred to the respective chapters in Refs.~\cite{Hecht,BornWolf,Roemer}, on which we have heavily drawn. 

\subsubsection{Experiment} \label{experiment}

{\itshape Law of reflection.}---Among all the facts adduced in the preceding subsections, only the law of reflection dates back to pre-modern times.
In fact, this law had traditionally been ascribed to Euclid (ca.~365-300 BC), in whose {\it Catoptrica} it can be found (see Refs.~\cite[p.~10]{Cajori} or \cite[p.~119]{Sarton}). 
Nowadays, however, this work is believed to be apocryphal (see ibid.~or \cite[p.~12]{DarrigolOptics}). In any case, the law of reflection was already known
to the ancients as it can also be found in the works of Archimedes (ca.~287-212 BC), Hero (floruit ca.~60 AD), and Ptolemy (ca.~100-170 AD) \cite[p.~13]{DarrigolOptics}. 
The fact that the incident and the reflected light rays lie in the same plane which also contains the surface normal has probably been known throughout
all this time, but explicitly, it can at best be traced back to the important Arab scholar Ibn Al Haitam (also: Al Hazen; ca.~965-1040 AD) \cite[pp.~21/22]{Cajori}.

{\itshape Law of refraction.}---Compared to the law of reflection, the law of refraction is much younger, dating back to W.~Snellius (1580 or 1591-1626) who discovered 
it in 1620 \cite[p.~227]{Simonyi}. The unpublished manuscript, however, is not extant. 
In its present form, the ``law of sines'' was given in 1637 by R.~Descartes (1596-1650) in {\it La Dioptrique} \cite[p.~83]{Cajori}.
Already in 1655, it appeared in the monumental treatise {\it De corpore} of the English philosopher-physicist T.~Hobbes (1588-1679).
Not much later, namely in 1666, I.~Newton (1643-1727) discovered the phenomenon of dispersion, which implies that actually each wavelength has its 
own index of refraction (see Refs.~\cite[p.~xxvi]{BornWolf} or \cite[p.~135]{Darmstaedter}). Thus, around 1700 AD the most basic facts related to refraction and reflection 
independently of possible polarization effects were well-known to the scientific community.

{\itshape Light polarization.}---Although the polarization of light as such had already been described by C.~Huygens (1629-1695) in 1690 \cite[pp.~2/3]{Roemer},
it was not until 1808/9 that E.-L.~Malus (1775-1812) observed that reflected rays may be polarized \cite[p.~xxvii]{BornWolf}, \cite[p.~37]{LaueHist}. In 1816/9, D.\,F.\,J.~Arago (1786-1853) together with the ingenious outsider A.-J.~Fresnel (1788-1827) then proved that light rays polarized orthogonally to each other do not interfere \cite[p.~xxviii]{BornWolf}, \cite[p.~37]{LaueHist}. 
Correspondingly, T.~Young (1773-1829) had hypothesized already in 1817 that light waves have to be transverse \cite[p.~xxviii]{BornWolf}, which fitted well into this picture.

{\itshape Discovery of Fresnel equations.}---In this situation, the polarization dependence of reflection and refraction had finally become a natural 
problem to study, and, astonishingly enough,
it was solved very soon, in 1823, by A.-J.~Fresnel \cite[p.~42]{BornWolf}. 
Fortunately, the equations now bearing his name could be corroborated
almost immediately after their formulation by the ensuing theoretical derivation of the so-called Brewster angle, 
which had been discovered independently by D.~Brewster (1781-1868) back in 1815 \cite[p.~12]{Roemer}. 

We note that throughout all these developments, the index of refraction was defined by angular measurements. 
The first measurements of the speed of light in materials have only been performed by A.\,H.\,L.~Fizeau (1819-1896), 
and later, A.\,A.~Michelson (1852-1931) found agreement with refractive index measurements \cite[p.~120]{Drude}.

\subsubsection{Theory} \label{theory}

{\itshape Electromagnetic field theory.}---Despite C.~Huygens' tentative derivation of refraction and reflection from field theoretical considerations \cite[p.~xxvi]{BornWolf},
by and large, the development of electromagnetic field theory went independently of the above experimental developments.
The distinction between the fields $\vec H$ and $\vec B$ was introduced by W.~Thomson (also: Lord Kelvin; 1824-1907) only in 1850. 
Later, these fields have been designated as ``magnetic force'' and ``magnetic induction'' by J.\,C.~Maxwell (1831-1879) \cite[p.~244]{Whittaker},
who successfully employed these concepts already in 1856 \cite[p.~90]{Longair}. 
The corresponding proportionality constant $\mu$ has been called ``permeability'' by W.~Thomson \cite[p.~245]{Whittaker}, after it had been 
introduced already in 1854 as a ``magnetization constant'' by M.~Faraday (1791-1867) \cite[p.~547]{Darmstaedter}. 
Moreover, Faraday already distinguished between para- and diamagnetism (see Refs.~\cite[p.~276]{Schreier} or \cite[p.~401]{Eliott}).
Similarly, the capacity enhancement by materials in a condenser was rediscovered---after unpublished results by H.~Cavendish (1731-1810)---by Faraday in 1837, thus leading to the
introduction of the ``specific inductive capacity'', nowadays called {\it permittivity} or {\it dielectric constant} \cite[p.~1]{Boettcher}.
Furthermore, Faraday already mentioned a state of ``polarization'' \cite[pp.~327 f.]{Eliott}. 
By contrast, the so-called ``displacement field'' $\vec D$  was only introduced by Maxwell in 1864 (see Refs.~\cite[p.~330]{Eliott} or \cite{DarrigolED}). 
This finally allowed for the definition of the permittivity as the ratio between the displacement field  and the electric field \cite[p.~1]{Boettcher}
in the sense of the equation $\vec D=\varepsilon\hh\vec E$.

{\itshape Maxwell relation.}---In this situation, when both the electric permittivity, the magnetic permeability and the refractive index had been defined inde\-{}pendently, Maxwell
finally derived in 1865 an electromagnetic wave equation in materials with the wave velocity $v=c/\sqrt{\varepsilon_{\rm r}\h\mu_{\rm r}}$ \cite[p.~104]{DarrigolMax}.
The corre\-{}sponding formula for the refractive index, $n=\sqrt{\varepsilon_{\rm r} \h \mu_{\rm r}}$\h, can already be found \linebreak in his original work, Ref.~\cite[Eq.~(80)]{Maxwell65}.
However, as it stood, the relation could not be verified with the data available at that time (see Refs.~\cite[vol.~2, p.~110]{Spasski} or \cite[p.~142]{Siegel}).
Astonishingly enough, the first tests were performed by Maxwell himself in 1871 for paraffin, but only approximate coincidence was found \cite[vol.~2, p.~110]{Spasski}.
Instead, it is L.\,E.~Boltzmann (1844-1906) who is credited with the verification of the Maxwell relation in 1874 \cite[p.~280]{Schreier}.
While Faraday had still assumed that the dielectric constants of all gases coincide \cite[p.~726]{Darmstaedter}, the differences in the
permittivities of gases were first measured by E.\,W.~v.~Siemens (1816-1892) in 1859 by means of a plate capacitor \cite[p.~593]{Darmstaedter}. 
Thus, in 1874 Boltzmann was in a position to verify the approximate relation $n^2=\varepsilon_{\rm r}$ for gases or, more generally, for those substances that do not display dispersion.
In fact, many substances called ``associating'' at that time did not fulfill the Maxwell relation 
if the refractive index was measured at optical frequencies, while the dielectric constant could at best be measured independently at much lower frequencies (as is naturally the case with measurements relying on the plate capacitor). 
Fortunately, however, measurements of the refractive index at much lower frequencies became possible with the discovery of microwaves by H.\,R.\,Hertz (1857-1894) in 1886. 
Correspondingly,  P.\,K.\,L.~Drude (1863-1906) showed in 1897 that the ``associating'' substances (i.e., the substances showing dispersion) indeed also fulfill the approximate 
Maxwell relation $n^2=\varepsilon_{\rm r}$ at low frequencies \cite[p.~2]{Boettcher}, thereby requiring again the assumption that the
relative magnetic permeability roughly equals~one,~i.e.~$\mu_{\rm r}=1$.

{\itshape Derivation of Fresnel equations.}---Finally, with the new field theory of electromagnetism, it remained to be shown that also the Fresnel equations could be recovered.
Independently from the enquiry of the refractive index, whose standard formula $n^2=\varepsilon_{\rm r} \h\mu_{\rm r}$
had been taken for granted, this problem was solved by H.\,A.~Lorentz (1853-1928). He rederived the 
Fresnel equations from Maxwell's theory in 1875 in his doctoral thesis \cite[p.~40]{LaueHist}.
However, the derivation by Lorentz, which soon became standard, also required the assumption $\mu_{\rm r}=1$ (see \S\,\ref{subsec:standardDeriv}).
Interestingly, Lorentz later also introduced the distinction between microscopic and macroscopic electrodynamics \cite{Lorentz1902,Lorentz1916}
and can therefore be considered the founding father of the Standard Approach to electrodynamics in media.

\subsubsection{Conclusion} \label{hist_conc}

Fresnel's equations are older than electromagnetic field theory. In fact, they have been {\it derived} independently of the Standard Approach to electrodynamics in media. Moreover, they have even
been {\it verified} independently of this Standard Approach since, for example, the Brewster angle had already been discovered prior
to Maxwell's theory of electromagnetism.

In their original form of the ``sine law'' and the ``tangent law'', Fresnel's equations express the electromagnetic  field amplitudes in terms of (incident and refracted) angles only (see Eqs.~\eqref{orig_Fresnel1}--\eqref{eq_FresnelTangentLaw}). As such, these equations do not even require knowledge of the refractive index, the sole optical material parameter known at the time of their discovery.  The refractive index on its side had already been known long since at that time. Therefore, the Fresnel equations constituted (and in fact still constitute)
{\it highly predictive} statements, which as such would never have been accepted if they were not essentially true
(with all possible ``limitations'' that any phenomenological law allows for). 
Furthermore, the Fresnel equations---be it in the form of the sine and tangent laws, or in the form involving the refractive index---hold
independently of the latter's relation to other material parameters such as the electric permittivity or the magnetic permeability.
Thus, the Fresnel equations are extremely trustworthy
and can even serve as a test case for any theoretical approach to electrodynamics in media.

By contrast, the picture is much more complicated when it comes to the Standard Approach to electrodynamics in media and its ensuing
derivations of the refractive index and the Fresnel equations. In fact, when Maxwell derived his formula for the refractive index, $n^2=\varepsilon_{\rm r}\h\mu_{\rm r}$\hh,
this {\it was} a predictive formula (at least in principle), 
because the material parameters $\varepsilon_{\rm r}$ and $\mu_{\rm r}$ had been defined (and partly even measured) independently of the refractive index.
Unfortunately though, in the original interpretation where these electromagnetic material properties can be interpreted as 
capacity and inductance ``enhancement factors'', the outcome of this formula is disastrous to say the least (take the example of water \cite[Table 3.2]{Hecht}).
In retrospect, this is not even surprising, as it had been clear already since Newton's time that the refractive index depends on the frequency
and can hence not be equal to frequency-independent material constants or products thereof (as the permittivity and permeability at Maxwell's time really were).

Correspondingly, it became gradually clear to the scientific community that the standard formula---if it was to hold at all---should hold when all involved quantities refer
to the {\itshape same} frequency. However, as direct measurements of the permittivity at optical frequencies are impossible even today,
it appeared equally impossible to verify the standard formula as a matter of principle. Empirically, though, it turned out that for some substances (typically gases)
the refractive index is roughly independent of the frequency, such that the standard relation for the refractive index could finally be verified in certain special cases
by na\"{i}vely (so to speak, against the spirit of the equation) comparing the {\it optical} refractive index with the root of the {\it static} permittivity.
For those substances whose dispersion could independently be established as weak, this comparison showed in fact a significant agreement. Later, it became even possible
to verify the approximate Maxwell relation $n^2(\omega)=\varepsilon_{\rm r}(\omega)$ at low frequencies, where {\it both} the refractive index {\it and} 
the dielectric function could be measured independently (at the {\it same} frequency $\omega$).
Curiously enough, this then somehow counted as a verification of the original Maxwell relation \mbox{$n^2=\varepsilon_{\rm r}\h\mu_{\rm r}$\hh,} while in actual fact
it had always been the approximate relation $n^2=\varepsilon_{\rm r}$ which has been corroborated.

In addition, it must be emphasized that
at {\it optical} frequencies, not even the relation $n^2 = \varepsilon_{\rm r}$ could ever be verified experimentally, since there the dielectric function turns out to be  inaccessible.
Instead, at optical frequencies, this formula is usually used as a {\it defining} equation which allows one to deduce the dielectric function from
optical measurements. Fortunately, at present, the Maxwell relation obtains nevertheless a certain predictive power because the dielectric function can be {\it calculated} independently,
namely by {\it ab initio methods}. However, if the thus obtained dielectric function disagrees with the experiment, it is not clear whether
this is due to the approximations inherent in these ab initio methods or rather to a failure of the Maxwell relation used on the experimental side for determining the dielectric function.
Correspondingly, the original equation $n^2=\varepsilon_{\rm r}\h\mu_{\rm r}$ becomes {\itshape a fortiori} completely untestable in the optical r\'{e}gime, as this would require an independent measurement of $n$, $\varepsilon_{\rm r}$ and $\mu_{\rm r}$ at the same optical frequency.

In any case, in deducing the dielectric function from optical measurements, the Fresnel equations become instrumental as these allow for the determination of the refractive index in the first place. This is all the more noteworthy, since the derivation of the Fresnel equations from the Standard Approach even {\it requires} the relation $n^2=\varepsilon_{\rm r}$\hh. By contrast, if the standard formula $n^2 = \varepsilon_{\rm r} \h \mu_{\rm r}$ was true, then according to the Standard Approach, the Fresnel equations would have to be replaced by the pseudo-Fresnel equations \eqref{eq_pseudoFresnel1}--\eqref{eq_pseudoFresnel4}. These would not involve the refractive index but the wave impedance, and hence the dielectric function could not be deduced in the way it is usually done.
Furthermore, Fresnel's sine and tangent laws as well as the formula for the Brewster angle would not even be true in this case, as we will further explain in the next subsection.

\subsection{Critique of standard derivation}\label{subsec:critique}

We have shown in \S\,\ref{subsec:standardDeriv} that the Standard Approach to electrodynamics in media is not suitable for reproducing the Fresnel equations directly. Instead, 
the corresponding derivation  ends up with Eqs.~\eqref{eq_pseudoFresnel1}--\eqref{eq_pseudoFresnel4}, which we refer to as ``pseudo-Fresnel equations''. As they stand, these equations differ grossly from the original Fresnel equations \eqref{eq_Fresnel1}--\eqref{eq_Fresnel4}, and the latter can only be recovered from the pseudo-Fresnel equations by means of the additional {\it arbitrary assumption} 
(see the discussion in Ref.~\cite{Refr}) that $\mu_{\rm r}= 1$ would hold for all materials (at least at optical frequencies). 
This assumption, which is apparently always introduced by sleight of hand, implies the Maxwell relation $n^2=\varepsilon_{\rm r}$ and hence the equality of the inverse wave impedance and the refractive index. Correspondingly, sometimes already Eqs.~\eqref{eq_pseudoFresnel1}--\eqref{eq_pseudoFresnel4} are called
``Fresnel equations'', which is, however, both factually and historically inaccurate.

We particularly stress that the allegedly more fundamental equation 
for the refractive index, $n^2 = \varepsilon_{\rm r} \h \mu_{\rm r}$\hh, is therefore {\it not consistent with Fresnel's equations,} as the latter can only be 
recovered from the Standard Approach if one assumes that $\mu_{\rm r} = 1$ and consequently $n=1/Z=\sqrt{\varepsilon_{\rm r}}$\h. 
On the other hand, if the arbitrary assumption $\mu_{\rm r}=1$ is dropped, then the pseudo-Fresnel equations as derived in the Standard Approach 
actually {\itshape contradict} the independently established sine and tangent laws, Eqs.~\eqref{eq_FresnelSineLaw} and \eqref{eq_FresnelTangentLaw},
as well as the formulae for the Brewster angle, Eq.~\eqref{eq_Brewster}, for the reflectance, Eq.~\eqref{eq_basisReflSpect},
and for ellipsometric measurements of the dielectric funtion, Eq.~\eqref{eq_ellipso}.

However, an experimental deviation from these latter formulae---which could theoretically be accounted for by the introduction of a nontrivial magnetic permeability at
optical frequencies---has apparently never been reported. We hence conclude that Eqs.~\eqref{eq_pseudoFresnel1}--\eqref{eq_pseudoFresnel4} as derived from the Standard Approach
have no experimental basis. In actual fact, those laws which have been verified experimentally are precisely the 
{\itshape original} Fresnel equations \eqref{eq_Fresnel1}--\eqref{eq_Fresnel4} and their ramifications (Brewster angle, reflectance formula, ellipsometry).

Astonishingly enough, this problematic state of affairs is not even restricted to the derivation of the Fresnel equations in the Standard Approach.
Quite to the contrary, exactly the same problem is encountered already in the derivation of the refractive index itself:
Here, the Standard Approach to electrodynamics in media leads to the equation $n^2=\varepsilon_{\rm r} \h \mu_{\rm r}$\hh,
while for practical purposes only the aforementioned approximation $n^2=\varepsilon_{\rm r}$ is used. Precisely as in the case of the
Fresnel equations, this is usually justified by the claim that $\mu_{\rm r} = 1$ would hold at optical frequencies. Thus, in close parallel to the Fresnel equations, the {\it original} equation for the refractive index derived within the Standard Approach has no experimental basis. In fact, the considerations of \S\,\ref{subsec:histReview} clearly show that it had been the experimental verification 
of the allegedly approximate relation $n^2=\varepsilon_{\rm r}$\hh, which in the first place led to the acceptance of Maxwell's theory for the refractive index. 

In view of this alarming absence of experimental confirmation, the authors of the present article have drawn in Ref.~\cite{Refr} the clearcut
conclusion that the allegedly approximate relation $n^2=\varepsilon_{\rm r}$ is actually the right formula for the refractive index,
whereas the formula $n^2=\varepsilon_{\rm r} \h \mu_{\rm r}$ is plainly wrong and {\it therefore} lacking experimental evidence. 
Apart from a number of general arguments against the standard formula for the refractive index \cite[\S\,3.2]{Refr},
this conclusion had been confirmed independently by the direct rederivation of $n^2=\varepsilon_{\rm r}$ from first principles \cite[Sct.~4]{Refr}
within the Functional Approach to electrodynamics of media \cite{ED1,ED2,EDOhm,Refr,EDLor,EDWave,EDFullGF,EffWW}. In the following,
we will show that the same phenomenon arises when it comes to the Fresnel equations:
a straightforward derivation within a microscopic approach to electrodynamics in media directly leads to the Fresnel equations \eqref{eq_Fresnel1}--\eqref{eq_Fresnel4}
{\it without} the necessity of assuming $\mu_{\rm r} = 1$.

\section{Functional Approach} \label{subsec:derivation}

\subsection{Introduction}

We now come to the question of how the Fresnel equations can be reproduced within the framework of microscopic electrodynamics of media.
For this purpose, we first note that the assumption $\mu_{\rm r} = 1$ can already be introduced at the level of the 
boundary conditions \eqref{eq_BCrepl1}--\eqref{eq_BCrepl4} used in the Standard Approach.
We then get the {\it new} boundary conditions 
\begin{align}
\vec n\cdot(\varepsilon_{\rm r, \hh 2}\h\vec E_2-\varepsilon_{\rm r, \hh 1}\h\vec E_1)  &=0\,, \label{eq_BCF1}\\[3pt]
\vec n\times(\vec E_2-\vec E_1) &=0\,, \label{eq_BCF2}\\[3pt]
\vec n\cdot(\vec B_2-\vec B_1)  &=0\,,\label{eq_BCF3}\\[3pt]
\vec n\times(\vec B_2-\vec B_1) &=0\,. \label{eq_BCF4}
\end{align}
Since the Fresnel equations are obtained from the Standard Approach by setting $\mu_{\rm r}=1$ at the end of the derivation, one could alternatively
perform the same calculation postulating $\mu_{\rm r}=1$ from the very outset, thus starting from the above boundary conditions \eqref{eq_BCF1}--\eqref{eq_BCF4}.

Furthermore, as the representation of light rays by plane waves does not constitute a peculiarity of the Standard Approach,
the problem of deriving the Fresnel equations in the Functional Approach actually simply boils down to the reproduction of the boundary conditions given by Eqs.~\eqref{eq_BCF1}--\eqref{eq_BCF4}.
In other words, if it can be shown that the Functional Approach directly leads to Eqs.~\eqref{eq_BCF1}--\eqref{eq_BCF4} {\it without}
setting $\mu_{\rm r}=1$, then the Fresnel equations \eqref{eq_Fresnel1}--\eqref{eq_Fresnel4} follow exactly in the same way
as the pseudo-Fresnel equations \eqref{eq_pseudoFresnel1}--\eqref{eq_pseudoFresnel4} follow in the Standard Approach.

Thus,  in order to derive the Fresnel equations within the Functional Approach, we only have to derive the boundary conditions \eqref{eq_BCF1}--\eqref{eq_BCF4}.
For this purpose, we first stress that the Functional Approach is a {\itshape microscopic} field theory
which does away with the distinction between field theories ``in matter'' and ``in vacuo''. Instead, within the Functional
Approach all electromagnetic fields are determined by their respective Maxwell equations, whose
form is independent of the presence of media. Therefore, if we want to describe refraction within the Functional Approach, we are confronted
with different solutions of the microscopic Maxwell equations, which are restricted to different space regions (e.g., to an upper and a lower half-space  in the case of a flat interface between two different materials). 
The question then simply is: which conditions on these solutions can be deduced directly from the microscopic Maxwell equations?
To this problem we now~turn.

\subsection{Microscopic boundary conditions}

Given two solutions $\{\vec E_i, \h \vec B_i\}$ ($i=1,2$) of the microscopic Maxwell 
equations with their respective sources $\{\rho_i, \h \vec j_i\}$, 
we consider the problem of \linebreak whether these solutions can be ``glued'' together. By this, we mean the following: We single out a certain volume $V\subset\mathbb{R}^3$
and assume one solution to hold in its interior, while the other solution should hold in the exterior.
In other words, we restrict the fields $\{\vec E_1,\vec B_1\}$ to $V$ and the fields $\{\vec E_2,\vec B_2\}$ to $\overline V \equiv \mathbb R^3\backslash V$, 
and analogously for the sources. The question now is: are the resulting fields $\{\vec E,\vec B\}$, which are ``glued'' together from the individual solutions according to
\begin{align}
\vec E(\vec x,t)&:=\chi_V(\vec x) \h \vec E_1(\vec x,t)+\chi_{\overline V}(\vec x) \h \vec E_2(\vec x,t)\,, \label{glue_1} \\[5pt]
\vec B(\vec x,t)&:=\chi_V(\vec x) \h \vec B_1(\vec x,t)+\chi_{\overline V}(\vec x) \h \vec B_2(\vec x,t)\,, \label{glue_2}
\end{align}
again solutions of the Maxwell equations with appropriate sources? In these equations, $\chi_V(\vec x)$ denotes the characteristic function of the volume $V$, i.e.,
\begin{equation}
 \chi_{V}(\vec x) = \left\{ \begin{array}{ll} 1 & \textnormal{if } \, \vec x \in V \,, \\[5pt]
 0 & \textnormal{if } \, \vec x \not\in V \,, \end{array} \right.
\end{equation}
while $\chi_{\overline V}(\vec x) = 1 - \chi_V(\vec x)$ is the characteristic function of $\overline V$.

To answer this question, we first note that the fields $\{\vec E_1,\vec B_1\}$ and $\{\vec E_2, \vec B_2\}$ separately fulfill the Maxwell equations in the interior of $V$ and $\overline V$, and hence the decisive point is only the behavior of the fields at the boundary $\partial V$. Consider, for example, Gauss' law. With the definition of the surface normal as
\begin{equation}
\vec n(\vec x):=\nabla\chi_{\overline V}(\vec x)=-\nabla\chi_V(\vec x)\,,
\end{equation}
we find from Eq.~\eqref{glue_1} that
\begin{equation} \label{add_surf_charge}
\begin{aligned}
\nabla\cdot\vec E(\vec x,t) & = \frac{1}{\varepsilon_0} \h ( \hh \chi_V(\vec x) \h \rho_1(\vec x,t) + \chi_{\overline V}(\vec x) \h \rho_2(\vec x,t) ) \\[5pt]
 & \quad \, + \vec n(\vec x) \cdot (\vec E_2(\vec x,t)-\vec E_1(\vec x,t))\,.
\end{aligned}
\end{equation}
Now, if we require the ``glued'' fields to fulfill again Maxwell's equations, then
the right-hand side of this equation must be interpreted as the corresponding charge density $\rho(\vec x,t)$ of the ``glued'' fields.
This shows that an additional surface charge density $\rho_{\partial V}(\vec x, t)$ arises, which is given by the last term in Eq.~\eqref{add_surf_charge}. This term is singular and not present in the original solutions of the Maxwell equations.
Similarly, a straightforward calculation using the remaining Maxwell equations yields further conditions on the field behavior at the boundary, which we summarize as follows:
\begin{align}
\vec n\cdot(\vec E_2-\vec E_1)&=\rho_{\partial V}/\varepsilon_0\,,\label{eq_bcE2}\\[3pt]
\vec n\times(\vec E_2-\vec E_1)&=0\,,\label{eq_bcE1}\\[3pt]
\vec n\cdot(\vec B_2-\vec B_1)&=0\,,\label{eq_bcB1}\\[3pt]
\vec n\times(\vec B_2-\vec B_1)&=\mu_0 \h \vec j_{\partial V}\label{eq_bcB2}\,.
\end{align}
This means, the ``glued''  fields given by Eqs.~\eqref{glue_1}--\eqref{glue_2} indeed fulfill the microscopic Maxwell equations but with the following sources:
\begin{align}
\rho(\vec x,t)&=\chi_V(\vec x)\h\rho_1(\vec x,t)+\chi_{\overline V}(\vec x)\h\rho_2(\vec x,t)+\rho_{\partial V}(\vec x,t)\,, \label{eq_newSource1}\\[5pt]
\vec j(\vec x,t)&=\chi_V(\vec x)\h\vec j_1(\vec x,t)+\chi_{\overline V}(\vec x,t)\h\vec j_2(\vec x)+\vec j_{\partial V}(\vec x,t)\,, \label{eq_newSource2}
\end{align}
where the surface charge and current densities are given by 
\begin{align}
 \rho_{\partial V}(\vec x,t)&=\varepsilon_0\h\vec n(\vec x)\cdot(\vec E_2(\vec x,t)-\vec E_1(\vec x,t))\,,\label{eq_rhoV}\\[5pt]
 \vec j_{\partial V}(\vec x,t)&=\mu_0^{-1}\h\vec n(\vec x)\times(\vec B_2(\vec x,t)-\vec B_1(\vec x,t))\,.\label{eq_jV}
\end{align}
The above formulae \eqref{eq_bcE2}--\eqref{eq_bcB2} constitute the {\itshape general} boundary conditions on the microscopic electromagnetic fields derived within the Functional Approach.
(They may be compared with their counterparts in the Standard Approach given by Eqs.~\eqref{eq_BCstand1}--\eqref{eq_BCstand4}.)
In particular, Eqs.~\eqref{eq_bcE2} and \eqref{eq_bcB2} are actually {\itshape defining equations} for the surface charge and current densities, 
whereas Eqs.~\eqref{eq_bcE1} and \eqref{eq_bcB1} are necessary {\it constraints}, meaning that only such fields can be glued together which fulfill
these equations. We further note that the general boundary conditions \eqref{eq_bcE2}--\eqref{eq_bcB2} follow directly from the microscopic Maxwell equations and therefore
hold independently of the presence of media. The most important application, however, is given by the field behaviour at the interface of different materials.

Before proceeding with the derivation of Fresnel's equations in the next subsection, 
we perform a consistency check on the results obtained so far. In fact, we have shown that ``gluing'' together two solutions of the Maxwell equations requires the introduction of new source fields given by Eqs.~\eqref{eq_newSource1}--\eqref{eq_newSource2}. Since the ``glued'' electric and magnetic fields satisfy the Maxwell equations with these new sources, the latter necessarily have to fulfill the continuity equation, i.e.,
\begin{equation}
\partial_t \hh \rho(\vec x,t)+\nabla\cdot\vec j(\vec x,t) = 0 \,. \label{eq_desired}
\end{equation}
The reason for this is that the continuity equation can be deduced directly from the Maxwell equations (more precisely, from Gauss' law and Amp\`{e}re's law with displacement current). As the consistency check, we now verify the above condition by an explicit calculation. First, the defining equations \eqref{eq_newSource1} and \eqref{eq_newSource2} imply that
\begin{equation}
\partial_t \hh \rho=\chi_V \h \partial_t \hh \rho_1 + \chi_{\overline V} \h\hh \partial_t \hh \rho_2+\partial_t \hh \rho_{\partial V}\,,
\end{equation}
as well as
\begin{equation}
\nabla\cdot\vec j=\chi_V \nabla\cdot\vec j_1 + \chi_{\overline V}\h \nabla\cdot\vec j_2 +\vec n\cdot(\vec j_2-\vec j_1)+\nabla\cdot\vec j_{\partial V}\,.
\end{equation}
Using the respective continuity equations for the original sources terms,
\begin{equation}
\partial_t \hh \rho_i+\nabla\cdot\vec j_i =0 \,,
\end{equation}
we then obtain the equality
\begin{equation}
\partial_t \hh \rho+ \nabla\cdot\vec j=\partial_t \hh \rho_{\partial V}+\vec n\cdot(\vec j_2-\vec j_1)+\nabla\cdot\vec j_{\partial V}\,. \label{eq_preCont}
\end{equation}
Now, the surface charge is given explicitly by Eq.~\eqref{eq_rhoV}, hence
\begin{equation} \label{zwischen_1}
 \partial_t \hh \rho_{\partial V} = \varepsilon_0 \h \vec n \cdot (\partial_t \vec E_2 - \partial_t \vec E_1) \,.
\end{equation}
Similarly, we evaluate the divergence of the surface current using Eq.~\eqref{eq_jV} and the vector identity
\begin{equation}
\nabla\cdot(\vec A\times\vec B)=\vec B\cdot(\nabla\times\vec A)-\vec A\cdot(\nabla\times\vec B)\,,
\end{equation}
whereby we obtain
\begin{align}
\nabla\cdot\vec j_{\partial V}=\mu_0^{-1}\h\big((\vec B_2-\vec B_1)\cdot(\nabla\times\vec n)-\vec n\cdot(\nabla \times \vec B_2- \nabla \times \vec B_1)\big)\,. \label{eq_intermed}
\end{align}
On the other hand, with the vanishing rotation of gradient fields, 
\begin{equation}
\nabla\times\vec n(\vec x)=\nabla\times(\nabla\chi_{\overline V}(\vec x))=0\,,
\end{equation}
and with Amp\`{e}re's law,
\begin{equation}
\nabla\times\vec B=\mu_0 \h \vec j+\mu_0 \h \varepsilon_0 \h \partial_t\vec E\,, 
\end{equation}
Eq.~\eqref{eq_intermed} reverts to
\begin{equation}
\nabla\cdot\vec j_{\partial V}=-\vec n\cdot\big((\vec j_2-\vec j_1)+\varepsilon_0 \hh (\partial_t\vec E_2-\partial_t\vec E_1)\big)\,.
\end{equation}
Furthermore, by Eq.~\eqref{zwischen_1}, this can be recast into
\begin{equation} \label{cont_recast}
\nabla\cdot\vec j_{\partial V}=-\vec n\cdot(\vec j_2-\vec j_1)-\partial_t \hh \rho_{\partial V}\,.
\end{equation}
Finally, plugging this formula into Eq.~\eqref{eq_preCont} yields the desired result, Eq. \eqref{eq_desired}. Thus, 
we have shown that the introduction of the additional surface charges and currents given by Eqs.~\eqref{eq_rhoV}--\eqref{eq_jV} is 
consistent with the continuity equation for the total sources.

\subsection{Derivation of Fresnel equations}

In the preceding subsection, we have derived the {\itshape general} boundary conditions on the microscopic electromagnetic fields, which are given by Eqs. \eqref{eq_bcE2}--\eqref{eq_bcB2}. In this section, we come to the derivation of the approximate boundary conditions \eqref{eq_BCF1}--\eqref{eq_BCF4}, which form the starting point for the derivation of the Fresnel equations. First, we see immediately that Eq.~\eqref{eq_bcE1} coincides with Eq.~\eqref{eq_BCF2}, and Eq.~\eqref{eq_bcB1} coincides with Eq.~\eqref{eq_BCF3} anyway. Furthermore, Eq.~\eqref{eq_BCF4} can be reproduced from Eq.~\eqref{eq_bcB2} by setting the surface current to zero, as it is also assumed in the Standard Approach.
It therefore only remains to derive the boundary condition \eqref{eq_BCF1} from the respective Eq.~\eqref{eq_bcE2}.
For this purpose, we set again the surface current to zero, $\vec j_{\partial V} =0$, as in the Standard Approach. By the continuity equation 
in the form of Eq.~\eqref{cont_recast}, this implies the additional condition
\begin{equation}
-\partial_t \hh \rho_{\partial V} = \vec n \cdot ( \hh \vec j_2 - \vec j_1) \,.
\end{equation}
Now,  with the definition of the surface charge density, Eq.~\eqref{eq_rhoV}, and with Ohm's law in terms of  the proper conductivity (see e.g.~Ref.~\cite[\S\,2.5]{Refr}), 
this translates into
\begin{equation}
 -\varepsilon_0 \h \vec n \cdot (\partial_t \vec E_2 - \partial_t \vec E_1) = \vec n \cdot \big( \hh \widetilde \sigma_2 \hh \vec E_2 - \widetilde \sigma_1 \vec E_1 \big) \,.
\end{equation}
By performing a Fourier tranformation with respect to time, which implies the replacement  $-\partial_t\vec E\mapsto \i\omega\vec E$, we obtain the equivalent formula
\begin{equation}
\vec n\cdot\left(\left(1-\frac{\widetilde\sigma_2(\omega)}{\i\omega\hh\varepsilon_0}\right) \! \vec E_2 -\left(1- \frac{\widetilde\sigma_1(\omega)}{\i\omega\hh\varepsilon_0}\right) \! \vec E_1\right)=0\,.
\end{equation}
Finally, substituting the proper conductivity in terms of the dielectric function by means of the standard relation (see Refs.~\cite[\S\,2.5]{Refr} and \cite{OptTens})
\begin{equation} \label{stand_rel}
\varepsilon_{\rm r, \hh eff}(\omega)=1-\frac{\widetilde\sigma(\omega)}{\i\omega\varepsilon_0}\,,
\end{equation}
yields precisely the desired boundary condition \eqref{eq_BCF1}. Thus, under the standard assumption of vanishing surface currents,
we have derived the approximate boundary conditions \eqref{eq_BCF1}--\eqref{eq_BCF4}, which in turn correspond to the boundary conditions \eqref{eq_BCrepl1}--\eqref{eq_BCrepl4}
used in the Standard Approach for deriving the Fresnel equations. The point is, however, that the {\itshape correct} boundary condition \eqref{eq_BCF4} is {\itshape automatically} obtained in the Functional Approach, whereas the same equation follows only formally from the standard boundary condition \eqref{eq_BCrepl4} by setting $\mu_{\rm r}=1$. Consequently, the Functional Approach also
leads directly to the correct Fresnel equations \eqref{eq_Fresnel1}--\eqref{eq_Fresnel4} without the detour of the pseudo-Fresnel equations \eqref{eq_pseudoFresnel1}--\eqref{eq_pseudoFresnel4}.
This concludes our derivation of Fresnel's equations from the Functional Approach.

\section{Conclusion}

We have subjected the standard derivation of Fresnel's equations to a systematic critique, which is based on the following main arguments:
\begin{enumerate}
\item The Standard Approach to electrodynamics in media, which relies on the formula $n=\sqrt{\varepsilon_{\rm r}\h\mu_{\rm r}}$ \h for the refractive index,
does actually not reproduce the real Fresnel equations \eqref{eq_Fresnel1}--\eqref{eq_Fresnel4} written terms of this refractive index. 
Instead, it leads to the ``pseudo-Fresnel equations'' \eqref{eq_pseudoFresnel1}--\eqref{eq_pseudoFresnel4} in terms of the inverse wave impedance $Z^{-1} = \sqrt{\varepsilon_{\rm r}/\mu_{\rm r}}$\h.
\item Only the real Fresnel equations \eqref{eq_Fresnel1}--\eqref{eq_Fresnel4} imply the empirically verified ``sine law'' and ``tangent law'' 
as well as the formulae for the Brewster angle and for the deduction of the dielectric function from reflectivity and ellipsometric measurements (see Sct.~\ref{Sec:FresnelEqus}).
\item Within the Standard Approach, this state of affairs can only be remedied by the arbitrary assumption $\mu_{\rm r}=1$, 
which implies the allegedly approximate relation $n=\sqrt{\varepsilon_{\rm r}}$\h. In particular, 
this arbitrary assumption also leads to the equality $n = Z^{-1}$, by which the pseudo-Fresnel equations can be transformed into the real Fresnel equations.
\item It has been shown independently by the authors of this article that the allegedly exact formula $n=\sqrt{\varepsilon_{\rm r}\h\mu_{\rm r}}$ is untenable \cite{Refr}. Instead, within modern microscopic approaches to electrodynamics in media, one directly finds the formula $n=\sqrt{\varepsilon_{\rm r}}$ \h for 
the refractive index without the necessity of postulating $\mu_{\rm r}=1$ (see Ref.~\cite[\S\,4.4]{Refr}).
\item Similarly, it has been shown in the present article that the Functional Approach to electrodynamics of media directly reproduces the original (i.e., real) Fresnel equations \eqref{eq_Fresnel1}--\eqref{eq_Fresnel4}
without the detour of the pseudo-Fresnel equations \eqref{eq_pseudoFresnel1}--\eqref{eq_pseudoFresnel4}.
\end{enumerate}

\smallskip \noindent
Furthermore, we have derived the new, microscopic boundary conditions for
the electromagnetic fields, Eqs.~\eqref{eq_bcE2}--\eqref{eq_bcB2}, which replace the older formulae \eqref{eq_BCstand1}--\eqref{eq_BCstand4} used in the Standard Approach. These new boundary 
conditions follow exclusively from the requirement that the electric and magnetic fields satisfy the microscopic Maxwell equations at the interface
between two materials. Therefore, these new boundary conditions are not restricted to ``macroscopically averaged fields'', and thus they are generally applicable in ab initio materials science.

\section*{Acknowledgments}
This research was supported by the DFG grant HO 2422/12-1 and by the DFG RTG 1995. R.\,S. thanks the Institute for Theoretical Physics at TU Bergakademie Freiberg for its hospitality.

\bibliographystyle{model1-num-names}
\bibliography{/net/home/lxtsfs1/tpc/schober/Ronald/masterbib}

\begin{thebibliography}{67}
\expandafter\ifx\csname natexlab\endcsname\relax\def\natexlab#1{#1}\fi
\providecommand{\url}[1]{\texttt{#1}}
\providecommand{\href}[2]{#2}
\providecommand{\path}[1]{#1}
\providecommand{\DOIprefix}{doi:}
\providecommand{\ArXivprefix}{arXiv:}
\providecommand{\URLprefix}{URL: }
\providecommand{\Pubmedprefix}{pmid:}
\providecommand{\doi}[1]{\href{http://dx.doi.org/#1}{\path{#1}}}
\providecommand{\Pubmed}[1]{\href{pmid:#1}{\path{#1}}}
\providecommand{\bibinfo}[2]{#2}
\ifx\xfnm\relax \def\xfnm[#1]{\unskip,\space#1}\fi
\bibitem[{Nozi\`{e}res and Pines(1958{\natexlab{a}})}]{NozieresPines2}
\bibinfo{author}{P.~Nozi\`{e}res}, \bibinfo{author}{D.~Pines},
\newblock \bibinfo{title}{{\itshape A dielectric formulation of the many body
  problem: application to the free electron gas}},
\newblock \bibinfo{journal}{Il Nuovo Cimento} \bibinfo{volume}{{\bfseries 9}}
  (\bibinfo{year}{1958}{\natexlab{a}}) \bibinfo{pages}{470}.
\bibitem[{Nozi\`{e}res and Pines(1958{\natexlab{b}})}]{NozieresPines}
\bibinfo{author}{P.~Nozi\`{e}res}, \bibinfo{author}{D.~Pines},
\newblock \bibinfo{title}{{\itshape Electron interaction in solids. General
  formulation}},
\newblock \bibinfo{journal}{Phys. Rev.} \bibinfo{volume}{{\bfseries 109}}
  (\bibinfo{year}{1958}{\natexlab{b}}) \bibinfo{pages}{741}.
\bibitem[{Lindhard(1954)}]{Lindhard}
\bibinfo{author}{J.~Lindhard},
\newblock \bibinfo{title}{{\itshape On the properties of a gas of charged
  particles}},
\newblock \bibinfo{journal}{Dan. Mat. Fys. Medd.} \bibinfo{volume}{{\bfseries
  28}} (\bibinfo{year}{1954}) \bibinfo{pages}{1}.
\bibitem[{Adler(1962)}]{Adler}
\bibinfo{author}{S.~L. Adler},
\newblock \bibinfo{title}{{\itshape Quantum theory of the dielectric constant
  in real solids}},
\newblock \bibinfo{journal}{Phys. Rev.} \bibinfo{volume}{{\bfseries 126}}
  (\bibinfo{year}{1962}) \bibinfo{pages}{413}.
\bibitem[{Wiser(1963)}]{Wiser}
\bibinfo{author}{N.~Wiser},
\newblock \bibinfo{title}{{\itshape Dielectric constant with local field
  effects included}},
\newblock \bibinfo{journal}{Phys. Rev.} \bibinfo{volume}{{\bfseries 129}}
  (\bibinfo{year}{1963}) \bibinfo{pages}{62}.
\bibitem[{Ehrenreich and Cohen(1959)}]{Ehrenreichcohen}
\bibinfo{author}{H.~Ehrenreich}, \bibinfo{author}{M.~H. Cohen},
\newblock \bibinfo{title}{{\itshape Self-consistent field approach to the
  many-electron problem}},
\newblock \bibinfo{journal}{Phys. Rev.} \bibinfo{volume}{{\bfseries 115}}
  (\bibinfo{year}{1959}) \bibinfo{pages}{786}.
\bibitem[{Hanke(1978)}]{Hanke}
\bibinfo{author}{W.~Hanke},
\newblock \bibinfo{title}{{\itshape Dielectric theory of elementary excitations
  in crystals}},
\newblock \bibinfo{journal}{Adv. Phys.} \bibinfo{volume}{{\bfseries 27}}
  (\bibinfo{year}{1978}) \bibinfo{pages}{287}.
\bibitem[{Strinati(1988)}]{Strinati}
\bibinfo{author}{G.~Strinati},
\newblock \bibinfo{title}{{\itshape Application of the Green's functions method
  to the study of optical properties of semiconductors}},
\newblock \bibinfo{journal}{La Rivista del Nuovo Cimento}
  \bibinfo{volume}{{\bfseries 11}} (\bibinfo{year}{1988}) \bibinfo{pages}{1}.
\bibitem[{Onida et~al.(2002)Onida, Reining, and Rubio}]{Onida}
\bibinfo{author}{G.~Onida}, \bibinfo{author}{L.~Reining},
  \bibinfo{author}{A.~Rubio},
\newblock \bibinfo{title}{{\itshape Electronic excitations: density-functional
  versus many-body Green's-function approaches}},
\newblock \bibinfo{journal}{Rev. Mod. Phys.} \bibinfo{volume}{{\bfseries 74}}
  (\bibinfo{year}{2002}) \bibinfo{pages}{601}.
\bibitem[{Trevisanutto et~al.(2013)Trevisanutto, Terentjevs, Constantin,
  Olevano, and Sala}]{Trevisanutto13}
\bibinfo{author}{P.~E. Trevisanutto}, \bibinfo{author}{A.~Terentjevs},
  \bibinfo{author}{L.~A. Constantin}, \bibinfo{author}{V.~Olevano},
  \bibinfo{author}{F.~D. Sala},
\newblock \bibinfo{title}{{\itshape Optical spectra of solids obtained by
  time-dependent density functional theory with the jellium-with-gap-model
  exchange-correlation kernel}},
\newblock \bibinfo{journal}{Phys. Rev. B} \bibinfo{volume}{{\bfseries 87}}
  (\bibinfo{year}{2013}) \bibinfo{pages}{205143}.
\bibitem[{Li et~al.(2015)Li, Chen, Huang, and Li}]{LiLi}
\bibinfo{author}{Y.~Li}, \bibinfo{author}{H.~Chen}, \bibinfo{author}{L.~Huang},
  \bibinfo{author}{J.~Li},
\newblock \bibinfo{title}{{\itshape Ab initio study of the dielectric and
  electronic properties of multilayer $\mathrm{GaS}$ films}},
\newblock \bibinfo{journal}{J. Phys. Chem. Lett.} \bibinfo{volume}{{\bfseries
  6}} (\bibinfo{year}{2015}) \bibinfo{pages}{1059}.
\bibitem[{Yang et~al.(2015)Yang, D'Archangel, Sundheimer, Tucker, Boreman, and
  Raschke}]{Yang15}
\bibinfo{author}{H.~U. Yang}, \bibinfo{author}{J.~D'Archangel},
  \bibinfo{author}{M.~L. Sundheimer}, \bibinfo{author}{E.~Tucker},
  \bibinfo{author}{G.~D. Boreman}, \bibinfo{author}{M.~B. Raschke},
\newblock \bibinfo{title}{{\itshape Optical dielectric function of silver}},
\newblock \bibinfo{journal}{Phys. Rev. B} \bibinfo{volume}{{\bfseries 91}}
  (\bibinfo{year}{2015}) \bibinfo{pages}{235137}.
\bibitem[{Arboleda et~al.(2016)Arboleda, Santill\'an, Herrera, Muraca, Schinca,
  and Scaffardi}]{Arbodela16}
\bibinfo{author}{D.~M. Arboleda}, \bibinfo{author}{J.~M.~J. Santill\'an},
  \bibinfo{author}{L.~J.~M. Herrera}, \bibinfo{author}{D.~Muraca},
  \bibinfo{author}{D.~C. Schinca}, \bibinfo{author}{L.~B. Scaffardi},
\newblock \bibinfo{title}{{\itshape Size-dependent complex dielectric function
  of $\mathrm{Ni}$, $\mathrm{Mo}$, $\mathrm{W}$, $\mathrm{Pb}$, $\mathrm{Zn}$
  and $\mathrm{Na}$ nanoparticles. Application to sizing}},
\newblock \bibinfo{journal}{J. Phys. D: Appl. Phys.}
  \bibinfo{volume}{{\bfseries 49}} (\bibinfo{year}{2016})
  \bibinfo{pages}{075302}.
\bibitem[{Bejaoui et~al.(2016)Bejaoui, Alonso, Garriga, Campoy-Quiles,
  {Go\~{n}i}, Hetsch, Kershaw, Rogach, To, Foo, and Zapien}]{Bejaoui16}
\bibinfo{author}{A.~Bejaoui}, \bibinfo{author}{M.~I. Alonso},
  \bibinfo{author}{M.~Garriga}, \bibinfo{author}{M.~Campoy-Quiles},
  \bibinfo{author}{A.~R. {Go\~{n}i}}, \bibinfo{author}{F.~Hetsch},
  \bibinfo{author}{S.~V. Kershaw}, \bibinfo{author}{A.~L. Rogach},
  \bibinfo{author}{C.~H. To}, \bibinfo{author}{Y.~Foo}, \bibinfo{author}{J.~A.
  Zapien},
\newblock \bibinfo{title}{{\itshape Evaluation of the dielectric function of
  colloidal $\mathrm{Cd_{1-x}Hg_{x}Te}$ quantum dot films by spectroscopic
  ellipsometry}},
\newblock \bibinfo{journal}{Appl. Surf. Sci.}  (\bibinfo{year}{2016}).
\bibitem[{Feneberg et~al.(2016)Feneberg, Nixdorf, Lidig, Goldhahn, Galazka,
  Bierwagen, and Speck}]{Feneberg16}
\bibinfo{author}{M.~Feneberg}, \bibinfo{author}{J.~Nixdorf},
  \bibinfo{author}{C.~Lidig}, \bibinfo{author}{R.~Goldhahn},
  \bibinfo{author}{Z.~Galazka}, \bibinfo{author}{O.~Bierwagen},
  \bibinfo{author}{J.~S. Speck},
\newblock \bibinfo{title}{{\itshape Many-electron effects on the dielectric
  function of cubic ${\mathrm{In}}_{2}{\mathrm{O}}_{3}$: effective electron
  mass, band nonparabolicity, band gap renormalization, and Burstein-Moss
  shift}},
\newblock \bibinfo{journal}{Phys. Rev. B} \bibinfo{volume}{{\bfseries 93}}
  (\bibinfo{year}{2016}) \bibinfo{pages}{045203}.
\bibitem[{Hassanien(2016)}]{Hassanien16diel}
\bibinfo{author}{A.~S. Hassanien},
\newblock \bibinfo{title}{{\itshape Studies on dielectric properties,
  opto-electrical parameters and electronic polarizability of thermally
  evaporated amorphous $\mathrm{Cd_{50}S_{50-x}Se_{x}}$ thin films}},
\newblock \bibinfo{journal}{J. Alloy. Comp.} \bibinfo{volume}{{\bfseries 671}}
  (\bibinfo{year}{2016}) \bibinfo{pages}{566}.
\bibitem[{Nuzhnyy et~al.(2016)Nuzhnyy, Petzelt, Borodavka, Van\v{e}k,
  \v{S}imek, Trunec, and Maca}]{Nuzhnyy16}
\bibinfo{author}{D.~Nuzhnyy}, \bibinfo{author}{J.~Petzelt},
  \bibinfo{author}{F.~Borodavka}, \bibinfo{author}{P.~Van\v{e}k},
  \bibinfo{author}{D.~\v{S}imek}, \bibinfo{author}{M.~Trunec},
  \bibinfo{author}{K.~Maca},
\newblock \bibinfo{title}{{\itshape Effective infrared reflectivity and
  dielectric function of polycrystalline alumina ceramics}},
\newblock \bibinfo{journal}{Phys. Status Solidi B}  (\bibinfo{year}{2016}).
\bibitem[{{M. Yu. Seyidov} et~al.(2016){M. Yu. Seyidov}, Mikailzade,
  Suleymanov, Bulut, and Salehli}]{Seyidov16}
\bibinfo{author}{{M. Yu. Seyidov}}, \bibinfo{author}{F.~A. Mikailzade},
  \bibinfo{author}{R.~A. Suleymanov}, \bibinfo{author}{N.~Bulut},
  \bibinfo{author}{F.~Salehli},
\newblock \bibinfo{title}{{\itshape The influence of uniaxial compressive
  stress on the phase transitions and dielectric properties of
  $\mathrm{NaNO_2}$}},
\newblock \bibinfo{journal}{J. Phys. Chem. Solids} \bibinfo{volume}{{\bfseries
  93}} (\bibinfo{year}{2016}) \bibinfo{pages}{22}.
\bibitem[{Vos and Grande(2017)}]{Vor17}
\bibinfo{author}{M.~Vos}, \bibinfo{author}{P.~L. Grande},
\newblock \bibinfo{title}{{\itshape Extracting the dielectric function from
  high-energy REELS measurements}},
\newblock \bibinfo{journal}{Surf. Interface Anal.}  (\bibinfo{year}{2017}).
\bibitem[{Zheng et~al.(2017)Zheng, Tao, and Rappe}]{Zheng17}
\bibinfo{author}{F.~Zheng}, \bibinfo{author}{J.~Tao}, \bibinfo{author}{A.~M.
  Rappe},
\newblock \bibinfo{title}{{\itshape Frequency-dependent dielectric function of
  semiconductors with application to physisorption}},
\newblock \bibinfo{journal}{Phys. Rev. B} \bibinfo{volume}{{\bfseries 95}}
  (\bibinfo{year}{2017}) \bibinfo{pages}{035203}.
\bibitem[{Resta and Vanderbilt(2007)}]{Resta07}
\bibinfo{author}{R.~Resta}, \bibinfo{author}{D.~Vanderbilt},
\newblock \bibinfo{title}{{\itshape Theory of polarization: a modern
  approach}},
\newblock in: \bibinfo{editor}{K.~M. Rabe}, \bibinfo{editor}{C.~H. Ahn},
  \bibinfo{editor}{J.~M. Triscone} (Eds.), \bibinfo{booktitle}{{\itshape
  Physics of ferroelectrics: a modern perspective}},
  \bibinfo{publisher}{Springer-Verlag}, \bibinfo{address}{Berlin/Heidelberg},
  \bibinfo{year}{2007}.
\bibitem[{Resta(2010)}]{Resta10}
\bibinfo{author}{R.~Resta},
\newblock \bibinfo{title}{{\itshape Electrical polarization and orbital
  magnetization: the modern theories}},
\newblock \bibinfo{journal}{J. Phys. Condens. Matter}
  \bibinfo{volume}{{\bfseries 22}} (\bibinfo{year}{2010})
  \bibinfo{pages}{123201}.
\bibitem[{Vanderbilt and Resta(2006)}]{Vanderbilt}
\bibinfo{author}{D.~Vanderbilt}, \bibinfo{author}{R.~Resta},
\newblock \bibinfo{title}{{\itshape Quantum electrostatics of insulators:
  polarization, Wannier functions, and electric fields}},
\newblock in: \bibinfo{editor}{S.~G. Louie}, \bibinfo{editor}{M.~L. Cohen}
  (Eds.), \bibinfo{booktitle}{{\itshape Conceptual foundations of materials: a
  standard model for ground- and excited-state properties}},
  \bibinfo{publisher}{Elsevier B.V.}, \bibinfo{address}{Amsterdam},
  \bibinfo{year}{2006}.
\bibitem[{Giuliani and Vignale(2005)}]{Giuliani}
\bibinfo{author}{G.~F. Giuliani}, \bibinfo{author}{G.~Vignale},
  \bibinfo{title}{{\itshape Quantum theory of the electron liquid}},
  \bibinfo{publisher}{Cambridge University Press},
  \bibinfo{address}{Cambridge}, \bibinfo{year}{2005}.
\bibitem[{Kohanoff(2006)}]{Kohanoff}
\bibinfo{author}{J.~Kohanoff}, \bibinfo{title}{{\itshape Electronic structure
  calculations for solids and molecules: theory and computational methods}},
  \bibinfo{publisher}{Cambridge University Press},
  \bibinfo{address}{Cambridge}, \bibinfo{year}{2006}.
\bibitem[{Martin(2008)}]{Martin}
\bibinfo{author}{R.~M. Martin}, \bibinfo{title}{{\itshape Electronic structure:
  basic theory and practical methods}}, \bibinfo{publisher}{Cambridge
  University Press}, \bibinfo{address}{Cambridge}, \bibinfo{year}{2008}.
\bibitem[{Bechstedt(2015)}]{Bechstedt}
\bibinfo{author}{F.~Bechstedt}, \bibinfo{title}{{\itshape Many-body approach to
  electronic excitations: concepts and applications}}, volume
  \bibinfo{volume}{181} of \textit{\bibinfo{series}{\textnormal{Springer Series
  in Solid-State Sciences}}}, \bibinfo{publisher}{Springer-Verlag},
  \bibinfo{address}{Berlin/Heidelberg}, \bibinfo{year}{2015}.
\bibitem[{Sch\"afer and Wegener(2002)}]{SchafWegener}
\bibinfo{author}{W.~Sch\"afer}, \bibinfo{author}{M.~Wegener},
  \bibinfo{title}{{\itshape Semiconductor optics and transport phenomena}},
  \textnormal{Advanced Texts in Physics}, \bibinfo{publisher}{Springer-Verlag},
  \bibinfo{address}{Berlin/Heidelberg}, \bibinfo{year}{2002}.
\bibitem[{Kaxiras(2003)}]{Kaxiras}
\bibinfo{author}{E.~Kaxiras}, \bibinfo{title}{{\itshape Atomic and electronic
  structure of solids}}, \bibinfo{publisher}{Cambridge University Press},
  \bibinfo{address}{Cambridge}, \bibinfo{year}{2003}.
\bibitem[{Martin and Rothen(2002)}]{MartinRothen}
\bibinfo{author}{P.~A. Martin}, \bibinfo{author}{F.~Rothen},
  \bibinfo{title}{{\itshape Many-body problems and quantum field theory: an
  introduction}}, \bibinfo{publisher}{Springer-Verlag},
  \bibinfo{address}{Berlin/Heidelberg}, \bibinfo{year}{2002}.
\bibitem[{Jackson(1999)}]{Jackson}
\bibinfo{author}{J.~D. Jackson}, \bibinfo{title}{{\itshape Classical
  electrodynamics}}, \bibinfo{edition}{3rd} ed., \bibinfo{publisher}{John Wiley
  \& Sons, Inc.}, \bibinfo{address}{Hoboken, NJ}, \bibinfo{year}{1999}.
\bibitem[{Griffiths(1999)}]{Griffiths}
\bibinfo{author}{D.~J. Griffiths}, \bibinfo{title}{{\itshape Introduction to
  electrodynamics}}, \bibinfo{edition}{3rd} ed.,
  \bibinfo{publisher}{Prentice-Hall, Inc.}, \bibinfo{address}{Upper Saddle
  River, NJ}, \bibinfo{year}{1999}.
\bibitem[{Landau and Lifshitz(1984)}]{Landau}
\bibinfo{author}{L.~D. Landau}, \bibinfo{author}{E.~M. Lifshitz},
  \bibinfo{title}{{\itshape Electrodynamics of continuous media}},
  volume~\bibinfo{volume}{8} of \textit{\bibinfo{series}{\textnormal{Course of
  Theoretical Physics}}}, \bibinfo{edition}{2nd} ed.,
  \bibinfo{publisher}{Pergamon Press Ltd.}, \bibinfo{address}{Oxford},
  \bibinfo{year}{1984}.
\bibitem[{Starke and Schober(2015)}]{ED1}
\bibinfo{author}{R.~Starke}, \bibinfo{author}{G.~A.~H. Schober},
\newblock \bibinfo{title}{{\itshape Functional Approach to electrodynamics of
  media}},
\newblock \bibinfo{journal}{Phot. Nano. Fund. Appl.}
  \bibinfo{volume}{{\bfseries 14}} (\bibinfo{year}{2015})
  \bibinfo{pages}{1--34}. \bibinfo{note}{{See also arXiv:1401.6800
  [cond-mat.mtrl-sci]}}.
\bibitem[{Starke and Schober(2016{\natexlab{a}})}]{ED2}
\bibinfo{author}{R.~Starke}, \bibinfo{author}{G.~A.~H. Schober},
  \bibinfo{title}{{\itshape Ab initio materials physics and microscopic
  electrodynamics of media}}, \bibinfo{howpublished}{arXiv:1606.00445
  [cond-mat.mtrl-sci]}, \bibinfo{year}{2016}{\natexlab{a}}.
\bibitem[{Starke and Schober(2016{\natexlab{b}})}]{EDOhm}
\bibinfo{author}{R.~Starke}, \bibinfo{author}{G.~A.~H. Schober},
\newblock \bibinfo{title}{{\itshape Relativistic covariance of {Ohm's} law}},
\newblock \bibinfo{journal}{Int. J. Mod. Phys. D} \bibinfo{volume}{{\bfseries
  25}} (\bibinfo{year}{2016}{\natexlab{b}}) \bibinfo{pages}{1640010}.
  \bibinfo{note}{{See also arXiv:1409.3723 [math-ph]}}.
\bibitem[{Starke and Schober(2017{\natexlab{a}})}]{Refr}
\bibinfo{author}{R.~Starke}, \bibinfo{author}{G.~A.~H. Schober},
\newblock \bibinfo{title}{{\itshape Microscopic theory of the refractive
  index}},
\newblock \bibinfo{journal}{Optik} \bibinfo{volume}{{\bfseries 140}}
  (\bibinfo{year}{2017}{\natexlab{a}}) \bibinfo{pages}{62}. \bibinfo{note}{{See
  also arXiv:1510.03404 [cond-mat.mtrl-sci]}}.
\bibitem[{Starke and Schober(2017{\natexlab{b}})}]{EDLor}
\bibinfo{author}{R.~Starke}, \bibinfo{author}{G.~A.~H. Schober},
\newblock \bibinfo{title}{{\itshape Covariant response theory and the boost
  transform of the dielectric tensor}},
\newblock \bibinfo{journal}{Int. J. Mod. Phys. D} \bibinfo{volume}{{\bfseries
  26}} (\bibinfo{year}{2017}{\natexlab{b}}) \bibinfo{pages}{1750163}.
  \bibinfo{note}{{See also arXiv:1702.06985 [physics.class-ph]}}.
\bibitem[{Starke and Schober(2017{\natexlab{c}})}]{EDWave}
\bibinfo{author}{R.~Starke}, \bibinfo{author}{G.~A.~H. Schober},
\newblock \bibinfo{title}{{\itshape Linear electromagnetic wave equations in
  materials}},
\newblock \bibinfo{journal}{Phot. Nano. Fund. Appl.}
  \bibinfo{volume}{{\bfseries 26}} (\bibinfo{year}{2017}{\natexlab{c}})
  \bibinfo{pages}{41}. \bibinfo{note}{{See also arXiv:1704.06615
  [cond-mat.mtrl-sci]}}.
\bibitem[{Schober and Starke(2017)}]{EDFullGF}
\bibinfo{author}{G.~A.~H. Schober}, \bibinfo{author}{R.~Starke},
  \bibinfo{title}{{\itshape General form of the full electromagnetic Green
  function in materials physics}}, \bibinfo{howpublished}{arXiv:1704.07594
  [physics.class-ph]}, \bibinfo{year}{2017}.
\bibitem[{Starke and Schober(2016)}]{EffWW}
\bibinfo{author}{R.~Starke}, \bibinfo{author}{G.~A.~H. Schober},
  \bibinfo{title}{{\itshape Response Theory of the electron-phonon coupling}},
  \bibinfo{howpublished}{arXiv:1606.00012 [cond-mat.mtrl-sci]},
  \bibinfo{year}{2016}.
\bibitem[{Yu and Cardona(2010)}]{Cardona}
\bibinfo{author}{P.~Y. Yu}, \bibinfo{author}{M.~Cardona},
  \bibinfo{title}{{\itshape Fundamentals of semiconductors: physics and
  materials properties}}, \textnormal{Graduate Texts in Physics},
  \bibinfo{edition}{4th} ed., \bibinfo{publisher}{Springer-Verlag},
  \bibinfo{address}{Berlin/Heidelberg}, \bibinfo{year}{2010}.
\bibitem[{Nolting(2007)}]{NoltingEdyn}
\bibinfo{author}{W.~Nolting}, \bibinfo{title}{{\itshape Grundkurs Theoretische
  Physik 3: Elektrodynamik}}, \bibinfo{edition}{8th} ed.,
  \bibinfo{publisher}{Springer-Verlag}, \bibinfo{address}{Berlin/Heidelberg},
  \bibinfo{year}{2007}.
\bibitem[{Fox(2010)}]{Fox}
\bibinfo{author}{M.~Fox}, \bibinfo{title}{{\itshape Optical properties of
  solids}}, \bibinfo{edition}{2nd} ed., \bibinfo{publisher}{Oxford University
  Press}, \bibinfo{address}{Oxford}, \bibinfo{year}{2010}.
\bibitem[{Hecht(2002)}]{Hecht}
\bibinfo{author}{E.~Hecht}, \bibinfo{title}{{\itshape Optics}},
  \bibinfo{edition}{4th} ed., \bibinfo{publisher}{Addison Wesley},
  \bibinfo{address}{San Francisco, CA}, \bibinfo{year}{2002}.
\bibitem[{Born and Wolf(1999)}]{BornWolf}
\bibinfo{author}{M.~Born}, \bibinfo{author}{E.~Wolf}, \bibinfo{title}{{\itshape
  Principles of optics: electromagnetic theory of propagation, interference and
  diffraction of light}}, \bibinfo{edition}{7th} ed.,
  \bibinfo{publisher}{Cambridge University Press},
  \bibinfo{address}{Cambridge}, \bibinfo{year}{1999}.
\bibitem[{R\"omer(2005)}]{Roemer}
\bibinfo{author}{H.~R\"omer}, \bibinfo{title}{{\itshape Theoretical optics: an
  introduction}}, \bibinfo{publisher}{Wiley-VCH Verlag GmbH {\&} Co. KGaA},
  \bibinfo{address}{Weinheim}, \bibinfo{year}{2005}.
\bibitem[{Cajori(1962)}]{Cajori}
\bibinfo{author}{F.~Cajori}, \bibinfo{title}{{\itshape A history of physics in
  its elementary branches including the evolution of physical laboratories}},
  \bibinfo{publisher}{Dover Publications, Inc.}, \bibinfo{address}{New York},
  \bibinfo{year}{1962}.
\bibitem[{Sarton(1987)}]{Sarton}
\bibinfo{author}{G.~Sarton}, \bibinfo{title}{{\itshape Hellenistic science and
  the culture in the last three centuries B.C.}}, \bibinfo{publisher}{Dover
  Publications, Inc.}, \bibinfo{address}{New York}, \bibinfo{year}{1987}.
\bibitem[{Darrigol(2012)}]{DarrigolOptics}
\bibinfo{author}{O.~Darrigol}, \bibinfo{title}{{\itshape A history of optics
  from greek antiquity to the nineteenth century}}, \bibinfo{publisher}{Oxford
  University Press}, \bibinfo{address}{Oxford}, \bibinfo{year}{2012}.
\bibitem[{Simonyi(2004)}]{Simonyi}
\bibinfo{author}{K.~Simonyi}, \bibinfo{title}{{\itshape Kulturgeschichte der
  Physik: von den Anf\"angen bis heute}}, \bibinfo{edition}{3rd} ed.,
  \bibinfo{publisher}{Wissenschaftlicher Verlag Harri Deutsch GmbH},
  \bibinfo{address}{Frankfurt am Main}, \bibinfo{year}{2004}.
\bibitem[{Darmstaedter(1908)}]{Darmstaedter}
\bibinfo{author}{L.~Darmstaedter}, \bibinfo{title}{{\itshape Handbuch zur
  Geschichte der Naturwissenschaften und der Technik. In chronologischer
  Darstellung}}, \bibinfo{publisher}{J.~Springer}, \bibinfo{address}{Berlin},
  \bibinfo{year}{1908}.
\bibitem[{Laue(1950)}]{LaueHist}
\bibinfo{author}{M.~v. Laue}, \bibinfo{title}{{\itshape History of physics}},
  \bibinfo{publisher}{Academic Press}, \bibinfo{address}{New York},
  \bibinfo{year}{1950}.
\bibitem[{Drude(1959)}]{Drude}
\bibinfo{author}{P.~Drude}, \bibinfo{title}{{\itshape The theory of optics}},
  \bibinfo{publisher}{Dover Publications, Inc.}, \bibinfo{address}{New York},
  \bibinfo{year}{1959}.
\bibitem[{Whittaker(1910)}]{Whittaker}
\bibinfo{author}{E.~Whittaker}, \bibinfo{title}{{\itshape A history of the
  theories of aether and electricity}}, \bibinfo{publisher}{Hodges, Figgis, \&
  Co., Ltd.}, \bibinfo{address}{Dublin}, \bibinfo{year}{1910}.
\bibitem[{Longair(2003)}]{Longair}
\bibinfo{author}{M.~Longair}, \bibinfo{title}{{\itshape Theoretical concepts in
  physics: an alternative view of theoretical reasoning in physics}},
  \bibinfo{edition}{2nd} ed., \bibinfo{publisher}{Cambridge University Press},
  \bibinfo{address}{Cambridge}, \bibinfo{year}{2003}.
\bibitem[{Schreier(1988)}]{Schreier}
\bibinfo{editor}{W.~Schreier} (Ed.), \bibinfo{title}{{\itshape Geschichte der
  Physik. Ein Abriss}}, \bibinfo{publisher}{Deutscher Verlag der
  Wissenschaften}, \bibinfo{address}{Berlin}, \bibinfo{year}{1988}.
\bibitem[{Eliott(1993)}]{Eliott}
\bibinfo{author}{R.~S. Eliott}, \bibinfo{title}{{\itshape Electromagnetics:
  history, theory, and applications}}, \bibinfo{publisher}{IEEE Press},
  \bibinfo{address}{New York}, \bibinfo{year}{1993}.
\bibitem[{B\"{o}ttcher(1973)}]{Boettcher}
\bibinfo{author}{C.~J.~F. B\"{o}ttcher}, \bibinfo{title}{{\itshape Theory of
  electric polarization}}, volume \bibinfo{volume}{1: {\itshape dielectrics in
  static fields}}, \bibinfo{edition}{2nd} ed., \bibinfo{publisher}{Elsevier
  Science Publishers B.V.}, \bibinfo{address}{Amsterdam}, \bibinfo{year}{1973}.
\bibitem[{Darrigol(2000)}]{DarrigolED}
\bibinfo{author}{O.~Darrigol}, \bibinfo{title}{{\itshape Electrodynamics from
  Amp\`{e}re to Einstein}}, \bibinfo{publisher}{Oxford University Press},
  \bibinfo{address}{Oxford}, \bibinfo{year}{2000}.
\bibitem[{Darrigol(2005)}]{DarrigolMax}
\bibinfo{author}{O.~Darrigol}, \bibinfo{title}{{\itshape Les \'{e}quations de
  Maxwell: de MacCullagh \`{a} Lorentz}}, \bibinfo{publisher}{\'{E}dition
  Belin}, \bibinfo{address}{Paris}, \bibinfo{year}{2005}.
\bibitem[{Maxwell(1865)}]{Maxwell65}
\bibinfo{author}{J.~C. Maxwell},
\newblock \bibinfo{title}{{\itshape A dynamical theory of the electromagnetic
  field}},
\newblock \bibinfo{journal}{Phil. Trans. R. Soc. Lond.}
  \bibinfo{volume}{{\bfseries 155}} (\bibinfo{year}{1865})
  \bibinfo{pages}{459}.
\bibitem[{Spasski(1963)}]{Spasski}
\bibinfo{author}{B.~I. Spasski}, \bibinfo{title}{{\itshape Istoria fisiki}},
  \bibinfo{publisher}{Izdatel'stvo Moskovskogo Universiteta},
  \bibinfo{address}{Moscow}, \bibinfo{year}{1963}.
\bibitem[{Siegel(1991)}]{Siegel}
\bibinfo{author}{D.~M. Siegel}, \bibinfo{title}{{\itshape Innovation in
  Maxwell's electromagnetic theory: molecular vortices, displacement current,
  and light}}, \bibinfo{publisher}{Cambridge University Press},
  \bibinfo{address}{Cambridge}, \bibinfo{year}{1991}.
\bibitem[{Lorentz(1902)}]{Lorentz1902}
\bibinfo{author}{H.~A. Lorentz},
\newblock \bibinfo{title}{{\itshape The fundamental equations for
  electromagnetic phenomena in ponderable bodies, deduced from the theory of
  electrons}},
\newblock \bibinfo{journal}{Proc. Roy. Acad. Amsterdam}
  \bibinfo{volume}{{\bfseries 5}} (\bibinfo{year}{1902}) \bibinfo{pages}{254}.
\bibitem[{Lorentz(1916)}]{Lorentz1916}
\bibinfo{author}{H.~A. Lorentz}, \bibinfo{title}{{\itshape The theory of
  electrons and its applications to the phenomena of light and radiant heat}},
  \bibinfo{edition}{2nd} ed., \bibinfo{publisher}{B.\,G.~Teubner},
  \bibinfo{address}{Leipzig}, \bibinfo{year}{1916}.
\bibitem[{Starke et~al.(2017)Starke, Schober, Wirnata, and Kortus}]{OptTens}
\bibinfo{author}{R.~Starke}, \bibinfo{author}{G.~A.~H. Schober},
  \bibinfo{author}{R.~Wirnata}, \bibinfo{author}{J.~Kortus},
  \bibinfo{title}{{\itshape Wavevector-dependent optical properties from
  wavevector-independent conductivity tensor}},
  \bibinfo{howpublished}{arXiv:1708.06330 [physics.optics]},
  \bibinfo{year}{2017}.

\end{thebibliography}

\end{document}